\documentclass[preprint,12pt]{elsarticle}

\usepackage{amssymb}
\usepackage{amsmath}
\usepackage{multirow}
\usepackage{caption}
\usepackage{booktabs}
\usepackage{graphics}
\usepackage{standalone}
\usepackage{bm}
\usepackage{xcolor}
\usepackage[hidelinks]{hyperref}
\usepackage[capitalize]{cleveref}
\crefname{equation}{}{}
\usepackage{siunitx}
\DeclareSIUnit{\pu}{p.u.}
\newcommand{\Rdim}[1]{\ensuremath{\mathbb{R}^{#1}}}

\journal{Electric Power Systems Research}

\begin{document}

\begin{frontmatter}

\title{Physics-Informed Neural Networks: a Plug and Play Integration into Power System Dynamic Simulations}

\author[1]{Ignasi Ventura Nadal}
\ead{ignad@dtu.dk}
\author[2,3]{Jochen Stiasny}
\ead{j.b.stiasny@tudelft.nl}
\author[1]{Spyros Chatzivasileiadis}
\ead{spchatz@dtu.dk}

\affiliation[1]{organization={Department of Wind and Energy Systems},
    addressline={Technical University of Denmark}, 
    city={Elektrovej},
    postcode={Kgs. Lyngby}, 
    state={2800},
    country={Denmark}}
    
\affiliation[2]{organization={Faculty of Electrical Engineering, Mathematics and Computer Science},
    addressline={TU Delft}, 
    city={Mekelweg 4},
    postcode={Delft}, 
    state={2628 CD},
    country={The Netherlands}}

\affiliation[3]{organization={Integrated Energy Systems},
    addressline={AIT Austrian Institute of Technology GmbH}, 
    city={Giefinggasse 4},
    postcode={Vienna}, 
    state={1210},
    country={Austria}}

\begin{abstract}
Time-domain simulations are crucial for ensuring power system stability and avoiding critical scenarios that could lead to blackouts.
The next-generation power systems require a significant increase in the computational cost and complexity of these simulations due to additional degrees of uncertainty, non-linearity and states. Physics-Informed Neural Networks (PINN) have been shown to accelerate single-component simulations by several orders of magnitude. However, their application to current time-domain simulation solvers has been particularly challenging since the system's dynamics depend on multiple components.
Using a new training formulation, this paper introduces the first natural step to integrate PINNs into multi-component time-domain simulations. We propose PINNs as an alternative to other classical numerical methods for individual components. Once trained, these neural networks approximate component dynamics more accurately for longer time steps. Formulated as an implicit and consistent method with the transient simulation workflow, PINNs speed up simulation time by significantly increasing the time steps used.
For explanation clarity, we demonstrate the training, integration, and simulation framework for several combinations of PINNs and numerical solution methods using the IEEE 9-bus system, although the method applies equally well to any power system size.

\end{abstract}

\end{frontmatter}



\section{Introduction}
 
Time-domain simulation is one of the principal power system tools facing major computational challenges. Ensuring system stability and avoiding critical scenarios is becoming more challenging and inefficient in a system whose dynamics only become faster and its uncertainty rapidly grows. The needed operating tools must provide millions of critical simulations and evaluations in a very short time and run almost simultaneously with the operation. The numerical integration currently used in all existing simulation tools is modular and numerically stable, but it rapidly becomes very computationally expensive when studying fast dynamics \cite{stott}. Although recent efforts have been made to improve the solvers' efficiency, they are still constrained by the fundamental theory \cite{daediff, daeimprov}. To achieve significant computational advantage for time-domain simulations, we need to rethink core modeling decisions.

Physics-Informed Neural Networks (PINNs), leveraging the underlying physical models during training, have been shown to effectively learn the solution to any given system of ordinary differential equations (ODEs), providing high-speed and sufficiently accurate results over long time steps \cite{pinnsraissi, resnets, sciml}. First proposed in \cite{lagaris}, a trained PINN offers an alternative to the classical numerical methods, yielding fast and accurate results without requiring very small time steps to ensure accuracy and stability. These characteristics make them extremely attractive to power system time-domain simulations. Ref.~\cite{pinnmisyris} introduced PINNs to power systems by modeling the dynamics described by the so-called swing equation of a single-machine infinite bus.

A new wave of research followed to try to incorporate the proven speed-ups of PINNs into time-domain simulations \cite{pinnreview}. This is not a trivial task, as they need to show an advantage over well-established numerical methods and decades-old theory. Probably, the most critical challenge for adopting PINNs in power system dynamics is to show how they can be integrated into a large dynamic system, capturing the dynamics or constraints imposed by neighboring interconnected components. The first approaches captured the whole system dynamics with one model, demonstrating simulation speed-ups of up to 10'000 times compared to conventional solvers \cite{pinn1jochen, pinndae}. While this speed-up proved a major development in the field, the high up-front training cost and poor generalisation made it an unfeasible implementation approach. To overcome these scalability barriers, the authors in \cite{pinnsim1} train a model for every dynamical component and later connect them through a root-finding algorithm, successfully defining a new simulator based on PINNs. Although this new approach provides significant simulation speed-ups, it faces barriers to adoption as it requires a new trained model for every component involved. The users cannot use their existing models and tools, for which they have spent significant resources over several years to develop. 

Besides replacing existing physical models to accelerate simulations (so-called ``forward problems'' \cite{pinnsraissi}), 
PINNs can also estimate unknown parameters of aggregated models or black box components ( so-called ``inverse problems'') \cite{pinnid1}, \cite{pinnid2}. While inverse problems are out of the scope of this paper, as we focus on forward applications, the methods we introduce here also enable the seamless integration of neural networks addressing inverse problems. 

Another method arising from the machine learning community to improve computational modeling is NeuralODEs \cite{node_neurips}. Instead of approximating the ODE solution, these methods directly capture the ODE function, keeping the same structure as classical integration schemes. Several works for multi-component dynamic simulations have been proposed using NeuralODEs \cite{feasibility_dae, mattnew}. While this method provides a straightforward and modular integration of NN-based models to existing transient simulation workflow, so far it has not been shown to have any drastic speed advantages.

This paper offers a new approach that combines PINNs' significant speed-ups with the integration and modularity capabilities of NeuralODEs: we formulate PINNs that capture the dynamic evolution of individual components considering the interaction with the system. In other words, we learn directly the solution of the ODEs for subsets of the network (or single components) while considering their interaction with the rest of the system. The proposed methodology enables a plug and play integration of PINNs into existing time-domain simulators, as its implementation is scalable and can be directly applied to established simulation workflows. PINNs become an alternative integration method that provides more accurate solutions over larger time step sizes for the relevant individual components and, through that, achieves significant simulation speed-ups.

To the best of our knowledge, this work is the first to propose PINNs that seamlessly integrate with classical integration methods to speed up simulations. Instead of replacing the current simulation framework, we propose PINNs as a new integration method to improve the state-of-the-art framework's performance. The contributions of this paper are as follows:
\begin{itemize}
    \item We introduce a novel Physics-Informed Neural Networks (PINNs) implicit formulation consistent with the numerical methods used in power system simulation software. This formulation captures the dynamics of individual components considering their interaction with the system.
    \item We integrate PINNs into a time-domain simulation solver and show that PINNs, when trained for larger time steps, can provide an alternative integration method that significantly increases simulation accuracy and allows for much larger time step sizes.
\end{itemize}
We make our code available online as the first version of a toolbox \cite{git} to foster the adoption of PINNs in the power systems community. 

The remainder of this paper is structured as follows. In  Sec.~\ref{sec2:concept}, we present the problem formulation and conceptually describe how PINNs can integrate into the transient simulation workflow. In  Sec.~\ref{sec3:methodology}, we describe the transient simulation workflow and present the PINN formulation as an accurate numerical integration method. Sec.~\ref{sec4:numstudy} shows the results of integrating PINNs into solvers for different multi-machine system simulations. The discussion and conclusions are offered in  Sec.~\ref{sec5:discussion} and Sec.~\ref{sec6:conclusions}, respectively.

\section{Problem Formulation}\label{sec2:concept}

This section introduces the problem formulation and describes conceptually how we integrate PINNs into the solution algorithm.

We assume a dynamical system described by the system of Differential-Algebraic Equations (DAEs) of the form
\begin{subequations}\label{daesystem}
\begin{align}
        \frac{d}{dt}x &= f(x, y)\\
        0 &= g(x, y),
\end{align}
\end{subequations}
with state variables $x(t)\in \Rdim{n}$, algebraic variables $y(t) \in \Rdim{m}$, the update function $f: \Rdim{n+m} \rightarrow \Rdim{n}$ and the algebraic relationship $g: \Rdim{n+m} \rightarrow \Rdim{m}$. We assume this semi-explicit form of DAEs to be of index 1, which requires that $\frac{\partial g}{\partial y}$ is non-singular \cite{daesystems}.

In order to solve this set of equations, we first need to approximate the evolution of the differential state from its current value $x_n$ to a future value $x_{n+1}$ a time step size of $h$ ahead
\begin{equation}\label{eq:state_integration}
    x_{n+h} = x_n + \int^{t_n + h}_{t_n} f(x,y) dt.
\end{equation}
In this paper, we focus on simultaneous approaches to solve the RMS-based simulations described by \eqref{daesystem}. Although partitioned algorithms can offer computational benefits, simultaneous approaches are better suited for capturing stiff dynamics. The approximation of the integration is expressed as an implicit algebraic relationship. We apply this algebraic relationship to \eqref{eq:state_integration} and as a consequence \eqref{daesystem} becomes a system of algebraic equations, which we can solve with a root-finding algorithm such as Newton-Raphson. We elaborate on these steps in Sec.~\ref{subsec:solvedaes} and \ref{subsec:numerics}.

The accuracy of approximating the integral with an algebraic relation will depend on the time step size $h$ and the system characteristics governed by $f$ and $g$. This work proposes to use PINNs for some of the algebraic approximations instead of numerical schemes, such as Runge-Kutta schemes, to improve the overall approximation accuracy and accelerate simulations.

\begin{figure}[!ht]
    \centering
    \includegraphics[width=0.95\linewidth]{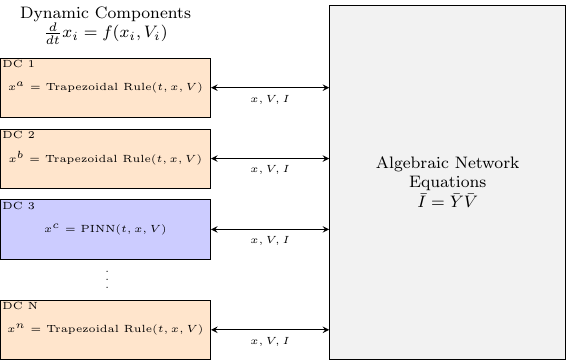}
    \caption{Depicted is a schematic of the power system dynamic simulations problem. The dynamic components are approximated by numerical schemes and solved together with the network equations. We introduce a PINN formulation (blue) that integrates modularly and compatibly with other methods, such as the trapezoidal rule (orange).}
    \label{fig:plugnplay}
\end{figure}

We introduce PINNs as a non-exclusive alternative to numerical schemes. PINNs provide higher accuracy over larger time steps than other methods. At the same time, they have less flexibility than other methods due to their upfront training cost. By introducing a PINN formulation consistent with other conventional schemes, we enable the integration of PINNs for individual components, see Fig.~\ref{fig:plugnplay}. Thus, we leverage the accuracy advantages of PINNs without needing to train for the whole system. We refer to this modular introduction of PINNs to simulations as a plug and play integration, as a trained PINN can be directly included in established transient simulation workflows. Sec.~\ref{sec4:numstudy} shows applications where the selective introduction of PINNs into simulations significantly increases accuracy and time step sizes. 

\textbf{Assumption 1:} We consider a system whose state and algebraic variables $x$ and $y$ can each be separated into two subsets denoted by $a$ and $b$, $x = \{x^a, x^b\}$ and $y = \{ y^a, y^b\}$. Based on this separation, the Jacobian of the dynamical system shows a block structure where the corresponding dynamics $f^a(x^a, y^a)$ and $f^b(x^b, y^a, y^b)$ can also be separated. This assumption holds for many networked systems such as power systems, where the dynamic states of the components are only coupled through algebraic variables. Hence, even for large power systems, the dynamic states can be separated into many subsets of low dimensionality. 

\begin{figure}[!ht]
    \centering
    \includegraphics[width=0.85\linewidth]{tikz_figures/sec2_matrix_structure.png}
    \caption{Block-structure of the Jacobian of the dynamical system by separating the variables and equations into two subsets, denoted by $a, b$. The filled areas indicate potentially non-zero elements in the matrix.}
    \label{fig:matrix_structure}
\end{figure}

Assumption 1 is illustrated in \Cref{fig:matrix_structure}. It allows us to approximate the integral of each subset in \cref{eq:state_integration} with different methods for $f^a$ and $f^b$. In this work, we apply a PINN for the integration of $f^a$
\begin{align}\label{eq:statea_integration}
    x^a_{n+1} = x^a_n + \int^{t_n + h}_{t_n} f^a(x^a,y^a) dt \approx x^a_n + \mathrm{PINN}(h, x^a,y^a),
\end{align}
which we present in \cref{subsect:pinnparam}. If $f^b$ shows itself a structure according to Assumption 1, the process can be repeated and another PINN be used.



\section{Methodology} \label{sec3:methodology}

This section first describes how the DAE in \Cref{daesystem} can be transformed into a system of algebraic equations and subsequently solved with the Newton-Raphson scheme. Then, we present how PINNs can be incorporated in this process, and lastly how the PINN is parameterized and trained. 

\subsection{Solving DAEs} \label{subsec:solvedaes}

The approach to addressing a system of DAEs is to first convert the differential equations into a system of algebraic equations and then solving it. The goal is the derivation of a set of algebraic equations 
\begin{align}\label{eq:dae_residual}
    F(x_{n+1}, y_{n+1}; h, x_{n}, y_{n}) = 0,
\end{align}
where $x_n, y_n$ and $x_{n+1}, y_{n+1}$ represent the system's state variable at the beginning and end of the time step of size $h$. Given the time step size $h$ and the initial condition $x_{n}, y_{n}$, we can solve for the values at the end of the time step $x_{n+1}, y_{n+1}$ such that $F(\cdot) = 0$. To that end, we use the Newton-Raphson algorithm which iteratively updates $X = \{x_{n+1}, y_{n+1}\}$ according to
\begin{equation} \label{newtonmethod}
    X^{(k+1)} = X^{(k)} - \bigg[\frac{\partial F}{\partial X} \bigg|_{X^{(k)}} \bigg]^{-1} F(X^{(k)}),
\end{equation}
where $k$ indicates the current iteration and $\frac{\partial F}{\partial X}$ is the Jacobian matrix of the residual function. The method is terminated when the updates reach a tolerance $\epsilon$ in $\max_i \big| X^{(k+1)}- X^{(k)}\big| < \epsilon$ or a maximum number of iterations $k^{\max}$. At this point, the values $x_{n+1}, y_{n+1}$ are stored as the solution and then used to solve for the following time step from $t_{n+1}$ to $t_{n+2}$.

\subsection{Approximating differential equations with Runge-Kutta schemes} \label{subsec:numerics}

The approximation of the differential equations with algebraic equations constitutes a critical step in the definition of the numerical method. Typically, implicit methods from the Runge-Kutta (RK) family are used, which assume a polynomial approximation to $x_{n+1}^a$. The simplest form of this family of methods is the trapezoidal rule, which leads to the following system of algebraic equations: 
\begin{subequations}
\begin{align}
    x_{n+1} &= x_{n} + \frac{h}{2} \left( f(x_n, y_n) + f(x_{n+1}, y_{n+1}) \right) \\
    0 &= g(x_{n+1}, y_{n+1}).
\end{align}
\end{subequations}
The resulting implicit equations form the residual function $F(x_{n+1}, y_{n+1})$ stated in \cref{eq:dae_residual}. Instead of the trapezoidal rule, other Runge-Kutta schemes can be used \cite{milano}. The choice of the scheme will determine the order of the local approximation error in dependence of the time step size $h$ and may have implications for the numerical stability of the solver. The choice becomes a trade-off between the improved accuracy and the longer run-times when increasing the order of the method. The trapezoidal rule is a common choice but requires small time step size $h$ to remain sufficiently accurate. This can lead to an overall slow simulation.



\subsection{Algebraize differential equations with PINNs}

As described in Sec.~\ref{sec2:concept}, we aim to perform the approximation of the integration step in \cref{eq:statea_integration} with a PINN
\begin{align}\label{eq:statea_integration_2}
    x^a_{n+1} = x^a_n + \int^{t_n + h}_{t_n} f^a(x^a,y^a) dt,
\end{align}
instead of a Runge-Kutta scheme such as the trapezoidal rule. The reason lies in the flexibility of PINNs to approximate general functions in an explicit manner. This allows for accurate approximations even for large time step sizes $h$. At the same time, the evaluation is very fast due to the PINN's explicit form. These benefits come at the price of requiring a learning process before the application.

An important step in training PINNs is the design of the inputs and the choice of the input domain. First, we select a range of time step sizes $h \in [0, h_{\max}]$. Second, we need to choose the domain $\mathcal{X}^a$ of the state $x^a$ such that all relevant system states are covered. Furthermore, it is desirable that $x^a_{n+1}$ also lies in $\mathcal{X}^a$ as it allows the repeated use of the same PINN, assuming that the underlying function $f^a$ does not vary with time. Lastly, we need to assume an evolution of $y^a(t)$ throughout the time step. We assume a linear evolution inside the time step, denoted by $\hat{y}^a(t)$
\begin{equation} \label{eq:assumedprofile}
    \hat{y}^a(t) = y^a_{n} + \frac{(y^a_{n+1}-y^a_{n})}{(t_{n+1}-t_n)} \; t,
\end{equation}
that is defined by the algebraic variables $y^a_{n}, y^a_{n+1}$ at the beginning and end of the time step. By replacing $y^a(t)$ in \cref{eq:statea_integration_2} with $\hat{y}^a(t)$, the integration can be determined and we learn an approximation $\hat{x}_{n+1}$ with a PINN 
\begin{equation}
    \hat{x}_{n+1}^a = x_n^a + h \, \mathrm{PINN}(h, x_n^a, y_n^a, y_{n+1}^a),
\end{equation}
that requires the inputs $h, x_n, y_n, y_{n+1}$. As all of these are readily included in $F(\cdot)$, this parametrization naturally fits the established solution approach for $F(\cdot) = 0$, iteratively converging to the final $\hat{x}_{n+1}$ and $y_{n+1}$. Within this Newton-Raphson algorithm, the computation of $\partial g(\hat{x}_{n+1}, \hat{y}_{n+1})/\partial y_{n+1}$ will be required. Thanks to Automatic Differentiation, these derivatives are easily computable in the PINN.

Given an input domain of the PINN that captures all the relevant operating conditions of the modeled dynamic component, we can evaluate the PINN recurrently as we would do with a RK method \cite{pinnsim1, rahul}. Such PINNs can support fixed and variable time step algorithms are supported as long as the time step is below $h_{max}$.

\begin{figure}[!ht]
\centering
\includegraphics[width=0.84\linewidth]{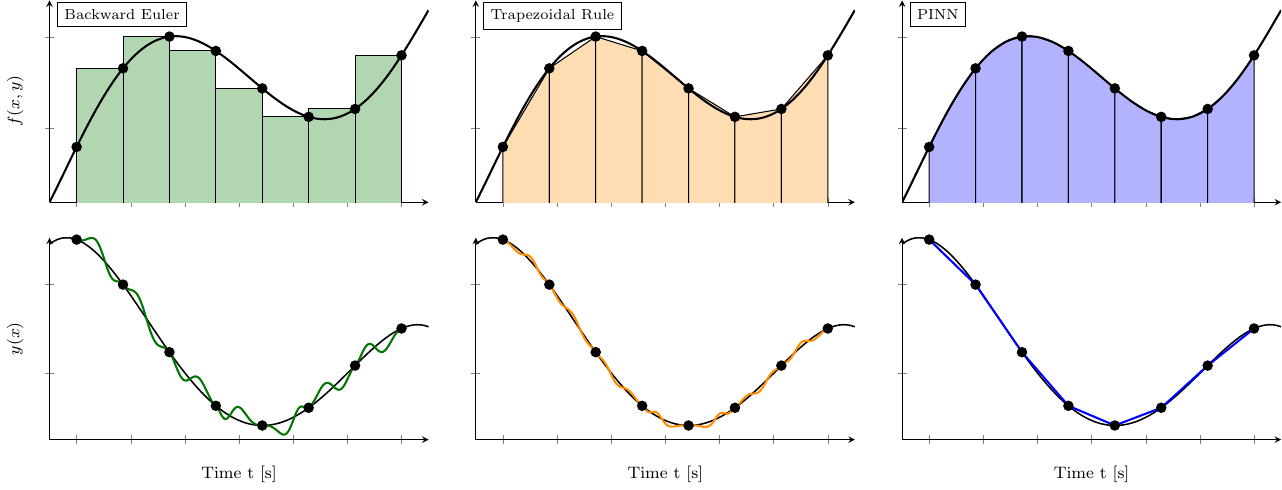}
\caption{Depicted is an approximate representation of a generic $f(x,y)$ and $y(x)$ evolution using the backward Euler, the trapezoidal rule and a PINN as the integration method. PINNs use the up-front training cost to capture better the studied trajectories for an accurate enough $y(x)$ evolution.}\label{fig:integrationschemes}
\end{figure}

\Cref{fig:integrationschemes} illustrates the essence of how PINNs work as numerical integration methods and why they can outperform other conventional methods. We compare how PINNs capture the trajectories against two of the most well-known implicit Runge-Kutta schemes, the backward Euler and the trapezoidal rule. Their main difference is how they approximate the integral of 8 the function $\int_{t_n}^{t_{n+h}} f^a(x_a, y_a) dt$ in \eqref{eq:statea_integration_2} to obtain the trajectory of xa. On the one hand, the backward Euler and the trapezoidal rule use rectangles and trapezoids to approximate the trajectory $x^a$. On the other hand, PINNs, as universal function approximators, learn an almost exact trajectory of $x^a$, given the evolution inside the time step of $y^a(x^a)$.

If the assumed $y(x)$ linear trajectory in the PINN fits the true one, PINNs are significantly more accurate than any other method for a time step size as large as it is trained for. However, $y(x)$ evolutions are rarely linear. The larger the time step size, the less accurate this linear approximation is. Thus, the time step sizes taken by PINNs are limited by the $y(x)$ approximation instead of the $f(x,y)$ one, which is very accurate. Sec.~\ref{sec4:numstudy} shows that, for a wide variety of time steps, the errors that arise from the PINNs' $y^a(x)$ approximation are smaller than the ones induced by the Runge-Kutta integral approximations. Thus, a solver that includes well-trained PINNs simulates significantly faster by performing longer time steps with the same accuracy.

\subsection{PINN Training and Setup} \label{subsect:pinnparam}

To parametrize a PINN we use a fully connected feed-forward NN with $K$ hidden layers and $N_k$ neurons. These hidden layers are parameterized by weight matrices $W^k$, bias vectors $b^k$ and a non-linear activation function $\sigma$. Thus, the mapping from inputs to outputs through the hidden layers is defined by 
\begin{equation}\label{eq:hiddenlayer}
    z_{k+1} = \sigma (W^{k+1} z_k + b^{k+1}), \; \forall k = 0, 1, ..., K-1.
\end{equation}
The output layer is adjusted according to \cite{lagaris} to enforce the initial conditions $x_n$ as hard constraints in the network's output. This approach improves the numerical consistency of our PINN. Thus, the output layer is stated as
\begin{equation} \label{eq:outputlayer}
    \hat{x}_{n+1} = x_n + h \big( \sigma(W^K z_K +b^K) \big).
\end{equation}

The training procedure to obtain the final weights and biases uses a loss function with two groups of losses, a data-based and a physics-based one. The data-based loss $\mathcal{L}_x$ 
\begin{equation}
    \mathcal{L}_x  = \frac{1}{N_x} \sum^{N_x}_{j=1} \left\Vert x^j_{n+1} - \Hat{x}^j_{n+1} \right\Vert_2^2, 
\end{equation}
uses the mean squared error to compare the PINN prediction to a dataset $\mathcal{D}_x$ of size $N_x$. This dataset consists of simulated results for different initial conditions $x_n$, time step sizes $h$ and evolutions of the algebraic equations parametrized by $y_n, y_{n+1}$
\begin{align}
    \mathcal{D}_x = \left\{ \left( h^i, x_n^i, y_n^i, y_{n+1}^i \right), x_{n+1}^i \right\}_{1 \leq i \leq N_x}.
\end{align}
Additionally, a physics-based loss $\mathcal{L}_c$
\begin{equation}
    \mathcal{L}_c = \frac{1}{N_c} \sum^{N_c}_{j=1} 
    \left\Vert \frac{d}{dt} \hat{x}(t^j) - f\left(\hat{x}^j, \hat{y}^j\right) \right\Vert_2^2,
\end{equation}
is evaluated on the dataset $\mathcal{D}_c$ of collocation points
\begin{align}
    \mathcal{D}_c = \left\{ \left( h^i, x_n^i, y_n^i, y_{n+1}^i \right) \right\}_{1 \leq i \leq N_c}.
\end{align}
$\mathcal{D}_c$ does not include $x_{n+1}^i$, as it can be generated without the need for simulations.

Both data-, and physics-based losses, inspired by the formulation proposed in \cite{pinn1jochen}, require the assumed linear profile for the constraining algebraic variables and the input time step to solve the differential equations.

The loss terms are added into the loss function with a hyper-parameter $\alpha$, so $\mathcal{L} = \mathcal{L}_x + \alpha \mathcal{L}_c$. This loss function will be minimized during the training by updating all the PINN's weights and biases of the architecture:
\begin{subequations}
\begin{align}
    \min_{\{W^k, b^k\}_{1\leq k\leq K}} \quad &\mathcal{L}_x(\mathcal{D}_x) + \alpha \mathcal{L}_c (\mathcal{D}_c)\\
    \text{s.t.}\quad & \cref{eq:hiddenlayer}, \cref{eq:outputlayer}.
\end{align}
\end{subequations}
The trained PINN can then be integrated into the DAE solver.

\section{Numerical Study} \label{sec4:numstudy}

In this section, we present test results that show how PINNs can replace and work together with numerical integration schemes to boost the simulation speed of power system dynamic simulations. We first define the study case and PINN implementation. We then present the numerical results of the presented plug and play PINN integration to simulation frameworks with the IEEE 9- and 57-bus system. It is important here to note that the method this paper introduces is general, and the results apply equally well to any power system size where the dynamic behavior of selected individual components is approximated by PINNs.

\subsection{Study case} \label{subsect:studycase}

We consider the current-balance form of a power system DAE \cite{sauerandpi}. Eqs.~\ref{difalgfinal} adapt the formulation of the general problem defined in \Cref{daesystem} to the power system as 
\begin{subequations} \label{difalgfinal}
    \begin{align}
        \frac{d}{dt} x_i(t) & = f_i(x_i, \Bar{V}_i), \; \forall i = 1, 2,..., N_c \label{eq:algindae} \\
        0 &= g(x, \Bar{V}_i), \label{eq:gform}
    \end{align}
\end{subequations}
where $x_i$ are the component states of the $i$-th component and $\Bar{V}_{i}$ are the terminal voltages at the $i$-th bus, which would compare to the $x$ and $y$ variables in \Cref{daesystem}, respectively.

The subsequent analysis uses the system formulation defined in Eqs.~\eqref{difalgfinal}. The state-space and algebraic models of the generators shown in \cref{eq:statespace} and \cref{eq:alggen} relate to \Cref{eq:algindae} and \Cref{eq:gform}, respectively. They represent the two-axis generator model as stated in \cite{sauerandpi}.
\begin{equation}\label{eq:statespace}
{
    \begin{bmatrix}
     T'_{do}  \\ 
     T'_{qo}   \\
     1 \\
     2H
    \end{bmatrix} \frac{d}{dt}
    \begin{bmatrix}
     E'_q \\
     E'_d \\
     \delta \\
     \Delta \omega
    \end{bmatrix} = 
    \begin{bmatrix}
     -E'_q - (X_d-X'_d)I_d +E_{fd}  \\ 
     -E'_d + (X_q-X'_q)I_q \\
     2\pi f \Delta \omega \\
     P_m-E'_dI_d-E'_qI_q-(X'_q-X'_d)I_dI_q - D \Delta \omega
    \end{bmatrix}
}
\end{equation}
\begin{equation} \label{eq:alggen}
    \begin{bmatrix}
        I_d \\ I_q
    \end{bmatrix} =
    \begin{bmatrix}
        R_s & -X'_q \\
        X'_d & R_s
    \end{bmatrix}^{-1}
    \begin{bmatrix}
        E'_d-V \sin (\delta - \theta) \\
        E'_q-V \cos (\delta - \theta)
    \end{bmatrix},
\end{equation}

where $\{E'_q, E'_d, \delta, \Delta \omega\}$ are the differential variables and $\{I_d, I_q\}$ the algebraic ones included in the formulation. For the following study, we assume a classical machine model by setting the reactances $X_q$ and $X'_q$ equal to $X'_d$ and assuming the internal voltages $E'_q$ and $E'_d$ to remain constant \cite{sauerandpi}. $P_m$ represents the mechanical power of the machine and $E_{fd}$ the excitation voltage. The variables $V$ and $\theta$ represent the terminal voltage. \Cref{eq:gform} considers Ohm's law across all the connected buses in the network, relating current to voltages. This formulation is developed in a per-unit system. The generator parameters and setpoints used in \eqref{eq:statespace} are displayed in the following \Cref{tab:machineparam}.

\begin{table}[!ht]
\centering
\captionsetup{skip=6pt}
\caption{Machine parameters and setpoints of the studied system in p.u. \cite{sauerandpi}.}\label{tab:machineparam} 
\begin{tabular}{cccccccc}
\toprule
\textbf{Machine} & \textbf{H} & \textbf{D} & \bm{$X_d$} & \bm{$X'_d$} & \bm{$R_s$} & \bm{$P_m$} & \bm{$E_{fd}$}  \\
\midrule
  1 & 23.64 & 2.364 & 0.146  & 0.0608 & 0.0 & 0.71  & 1.08 \\
  2 & 6.4   & 1.28  & 0.8958 & 0.1969 & 0.0 & 1.612 & 1.32 \\
  3 & 3.01  & 0.903 & 1.3125 & 0.1813 & 0.0 & 0.859 & 1.04 \\
  \bottomrule
\end{tabular}
\end{table}

\subsection{PINN Implementation}
Both the power system simulations and PINN training leverage the PyTorch library \cite{pytorch} with its automatic differentiation capabilities. We train the PINNs over a specified input range until reaching sufficient accuracy. We use the Adam optimizer for $3 \cdot 10^6$ epochs, using a delayed exponential decay function to decrease the learning rate. The training takes approximately four hours in an NVIDIA A100 GPU, hosted at \cite{hpc}. The PINNs consist of 64 neurons for each of the three hidden layers, using a $tanh$ activation function. We highlight that as a companion to this paper, we publish our code online and provide a detailed overview to integrate PINNs into time-domain simulations, see \cite{git}. To create the ground truth for our analyses, we use the simulation package Assimulo \cite{assimulo} to perform the DAE numerical simulations with a very small time step. All PINNs parametrize the time step size, and differential and algebraic variables within the ranges described in \Cref{tab:ranges_input}. All the initial conditions used in the subsequent numerical results are randomly sampled from the input domain to make sure we avoid model bias and that they perform accurately across all stated ranges.

\begin{table}[!ht]
\centering
\captionsetup{skip=6pt}
\caption{Input domain of initial conditions.}\label{tab:ranges_input}
\begin{tabular}{ccccc}
\toprule
\bm{$\Delta t$} & \bm{$\delta_{1,2,3}-\theta_{1,2,3}$} & \bm{$\Delta \omega_{1,2,3}$} & \bm{$V^{0,\Delta t}_{1,2,3}$} & \bm{$\Theta_{1,2,3}$} \\
\midrule
$[1, 40]$ ms & $[0, \frac{\pi}{3}]$ rad& $[-0.9, 0.9]$ Hz & $[0.97, 1.03]$ p.u. & $[-\pi, \pi]$ rad \\
\bottomrule
\end{tabular}
\end{table}

\subsection{Numerical Results}
The performance analysis of including PINNs in simulating frameworks is divided into two parts. The first part introduces a PINN into the simulation of a 9-bus system, showcasing conceptually how PINNs integrate and improve the dynamic simulations' performance. The second part scales the methodology to the 57-bus system, focusing on how several PINNs can seamlessly integrate into dynamic simulations and boost their performance.

In both cases, we study the performance of a DAE solver where the dynamics of one or more components are captured by a PINN and the rest by the conventional trapezoidal rule; we will call this the hybrid trapezoidal solver. We then benchmark the performance of the hybrid trapezoidal solver with a pure trapezoidal one, where all the generators are modeled with the trapezoidal rule. As one of the most well-known and used A-stable implicit Runge-Kutta schemes, the trapezoidal rule provides a solid reference method for comparison in this numerical analysis. Both solvers use exactly the same algorithm and only the integration method of one component changes. This helps us perform as objective a comparison as possible in evaluating both solvers' accuracy for different time step sizes. We use the same PINN to capture the dynamics of the replaced generators for both systems and for all scenarios.

\subsubsection{PINN Integration: A Conceptual and Computational Perspective }
In this study, we replace Generator 3 with a PINN, as described in \Cref{tab:solvers}. This generator presents the smallest inertia and damping of the system, shown in \Cref{tab:machineparam}. Thus, it presents the smallest time constant of the DAE. By replacing the trapezoidal rule with a trained PINN for Generator 3, we accurately capture the simulation's most determining dynamics for larger time steps. As shown in the subsequent results, learning the most critical dynamics enables larger and more accurate time steps in the system's simulation, reducing the number of global integrating time steps required.

\begin{table}[!ht]
\centering
\small
\captionsetup{skip=6pt}
\caption{Numerical integration method used for each of the system's generators.}\label{tab:solvers}
\renewcommand{\arraystretch}{1.25}
\begin{tabular}{c||c|c|c}
\textbf{Solver type} & \textbf{Generator 1} & \textbf{Generator 2} & \textbf{Generator 3}  \\
\hline
Hybrid solver & Trapezoidal rule& Trapezoidal rule & PINN \\
Pure solver & Trapezoidal rule& Trapezoidal rule & Trapezoidal rule \\
\end{tabular}
\end{table}

We analyze both global and local errors to show the true strengths and weaknesses of PINNs as a numerical technique. For presentation's clarity, we will focus on two variables: (i) the load angle of Generator 3, $\delta'_3 = \delta_3 - \theta_3$, and (ii) the terminal voltage magnitude of Generator 3, $V_3$. For insights into other variables, we post results from the remaining variables in the public toolbox \cite{git}. The errors shown depict the difference between the true and predicted values for each variable considered.

\begin{figure}[!ht]
\centering
\includegraphics[width=\linewidth]{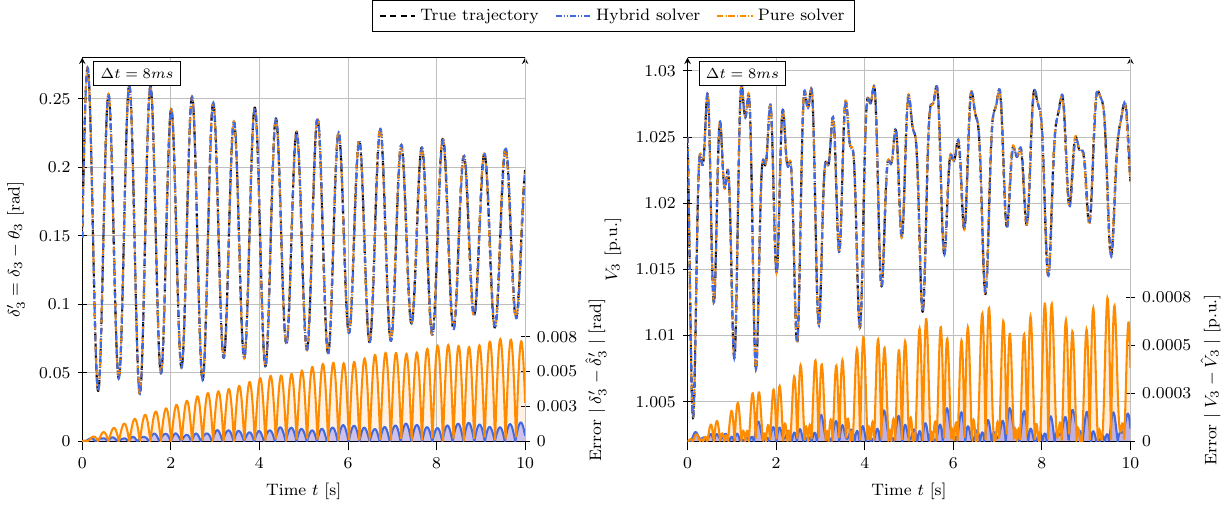}
\caption{Load angle and terminal voltage evolution over a 10-second trajectory with a time step of $\Delta t = 8 \, \textrm{ms}$. The left axis illustrates the true and predicted trajectories, while the right axis shows how the absolute error between the true and predicted ones evolves.}\label{fig:timesteps8}
\end{figure}

\noindent \textbf{1) Global error performance study} \\
\indent
We present the global error of both the pure and hybrid solvers over a ten-second simulation after a sudden decrease of mechanical torque in generator 2. The global errors result from the local truncation errors, which will be studied in the next part, and illustrate how these errors propagate over multiple time steps. They offer insights both on their long-term accuracy and numerical stability of the solvers. \Cref{fig:timesteps8} depicts the solvers' performance with a relatively small time step size of $\Delta t = 8 \, \textrm{ms}$. The trajectories follow the true trajectory very closely, and the errors are both within the acceptable range, i.e. they are all below the selected accuracy tolerance. However, after ten seconds, the hybrid solver errors are 5x-10x smaller, as they do not propagate as fast as the ones from the pure trapezoidal one. And this is already achieved by substituting only one out of the 3 machines with a PINN.

\begin{figure}[!ht]
\centering
\includegraphics[width=\linewidth]{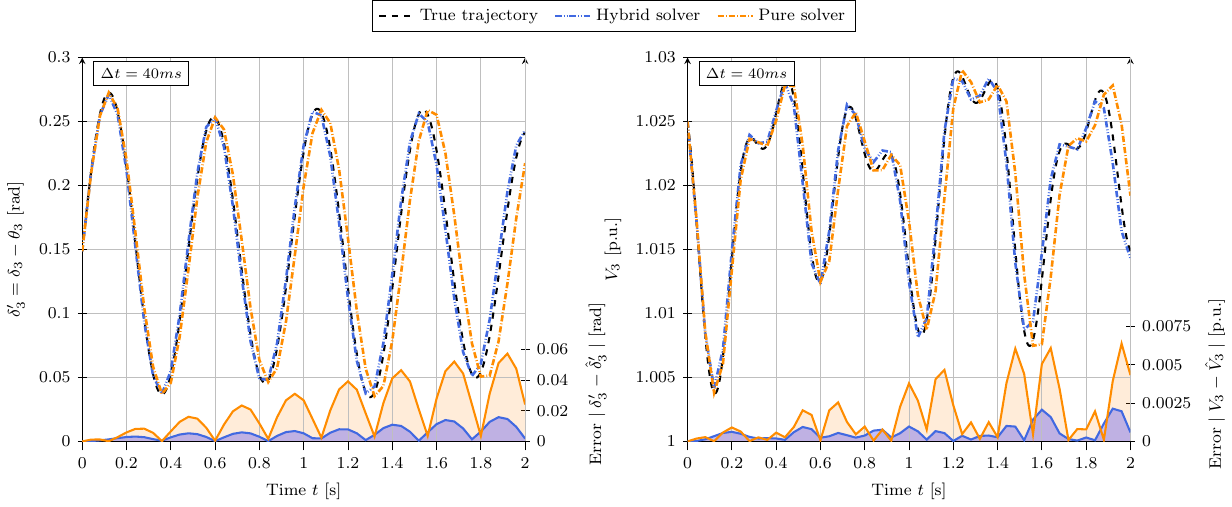}
\caption{Load angle and terminal voltage evolution over a 2-second trajectory with a time step $\Delta t = 40 \, \textrm{ms}$. The left axis illustrates the true and predicted trajectories, while the right axis shows how the absolute error between the true and predicted ones evolves}\label{fig:timesteps4}
\end{figure}

A very illustrating test for both solvers is to increase the time step size used in the simulation to a point where the predicted trajectory shows large mismatches from the true one. This phenomenon can be seen in \Cref{fig:timesteps4}, where the time step used is five times larger, i.e. $\Delta t = 40 \, \textrm{ms}$. In this case, the trajectories do not overlap. The predicted trajectories diverge from the true one as the errors accumulate over the two seconds. Increasing the time step sizes entails larger errors in both cases; however, they grow significantly faster with the pure solver than with the hybrid one.

\begin{figure}[!ht]
\centering
\includegraphics[width=0.85\linewidth]{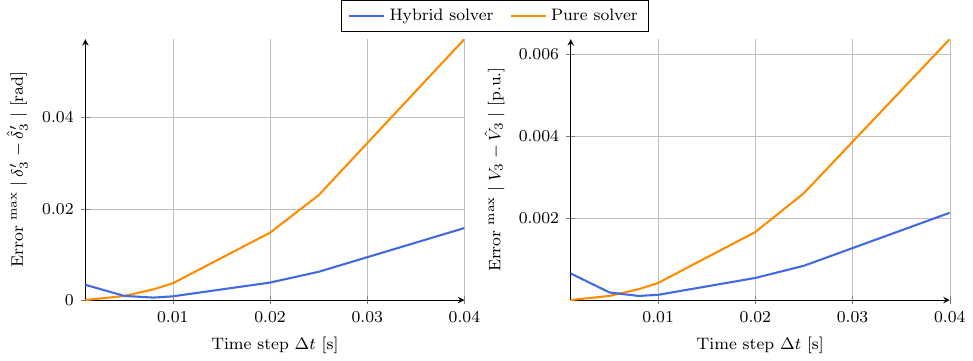}
\caption{Maximum errors obtained using different time step sizes over a 2-second simulation. The blue and orange trajectories depict the pure and hybrid solver, respectively.}\label{fig:maximums_evolution}
\end{figure}

\Cref{fig:maximums_evolution} presents how the errors of both studied solvers evolve for different time step sizes $\Delta t \in [1, 40] \, \textrm{ms}$. This illustration shows the simulation error in the load angle and terminal voltage magnitude of Bus 3 for the pure and hybrid solvers. The pure solver presents the characteristic curve of Runge-Kutta schemes, where the error increases exponentially as the simulation time step size increases. The interesting fact of this figure, however, is to see how the hybrid solver errors compare. For very small time step sizes, the PINN method cannot compete with the trapezoidal solver, which would asymptotically converge to zero as the time step sizes tend to zero. The key advantage comes as we increase the time step size. We can see that while the error of the pure solver evolves exponentially, the hybrid solver's error evolves more stably and smoothly. Thus, while PINNs do require an up-front training cost to learn the dynamics, this can be used as an advantage to then simulate dynamics several times faster.

\begin{figure}[!ht]
\centering
\includegraphics[width=0.85\linewidth]{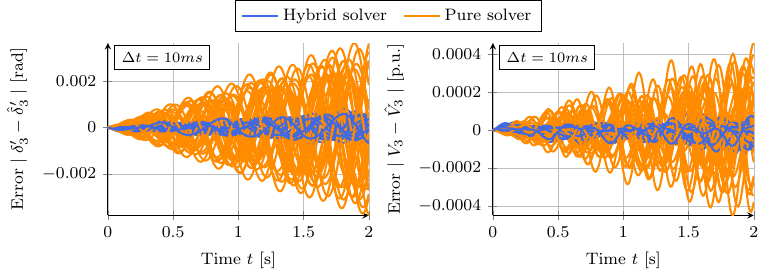}
\caption{Error evolutions for 30 different random initial conditions over a two-second simulation with a time step $\Delta t = 10 \, \textrm{ms}$. Each line depicts the error evolution of one simulation. The blue and orange colors represent the hybrid and pure solver, respectively.}\label{fig:30plots}
\end{figure}

\Cref{fig:30plots} highlights the multi-step errors for 30 random simulation initial conditions. Specifically, it depicts the error evolutions of both the pure and hybrid solver over two-second simulations. The pure solver errors propagate fast and steady as the iterations advance. However, by including a PINN in the simulation, the errors become more accurate and bounded, significantly outperforming the conventional simulation algorithms.

\noindent \textbf{2) Local error performance study} \\
\indent
Error propagation always starts from the local error. Thus, we continue the numerical analysis of the hybrid solver by isolating the error characteristics over a single time step. By looking at one time step, we can standardize the comparison between solvers. 

\begin{figure}[!ht]
\centering
\includegraphics[width=0.9\linewidth]{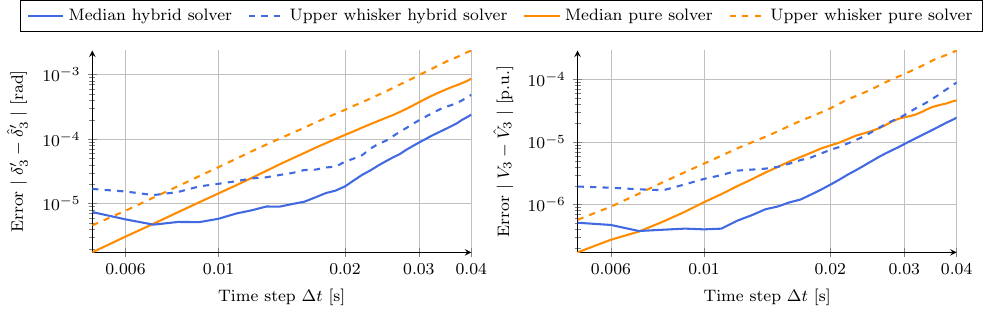}
\caption{Error distributions for 100 different random initial conditions of the pure and hybrid solvers. The solid and dashed lines depict the median and the upper whisker of the distributions in \emph{logarithmic} axes.}\label{fig:onetimestepevolution}
\end{figure}

\Cref{fig:onetimestepevolution} depicts the error distributions after one time step of a hundred different initial conditions randomly sampled from \Cref{tab:ranges_input}. For time steps $\Delta t \in [5, 40] \, \textrm{ms}$, we compute the median, interquartile range and the upper whisker of these distributions as for boxplot visualizations. The solid and dashed lines illustrate the distribution's median and upper whisker, respectively. While the pure solver increases its errors exponentially, i.e. linearly in the logarithmic scale, the hybrid solver presents a delayed exponential behaviour. Over a range of small time step sizes, the accuracy remains constant, and as the time step sizes become larger, the errors also start to grow exponentially. The smaller errors we observe over one time step with the hybrid solver depict the superiority of using a trained PINN as an integration method for individual components. This behaviour is seen across all variables, differential and algebraic, with the only exception of $\Delta \omega$, whose accuracy over a single time step remains the same.

\subsubsection{Scalability to larger systems with further PINN integration }
The introduced methodology enables seamless integration of PINNs, and in general, any NN, into power system dynamic simulations. This integration is not limited to the presented 9-bus system or one dynamic component. The plug and play integration framework can scale to any power system size and include several dynamic components captured by a PINN. In this section, we demonstrate the methodology using the IEEE 57-bus test system \cite{illinois}. This system consists of 57 buses, 7 generators, and 42 loads. The generators in buses 2, 3, 8, and 12 are set to the same parameters shown in \Cref{tab:machineparam} and replaced by the same PINN.

\begin{figure}[!ht]
\centering
\includegraphics[width=0.95\linewidth]{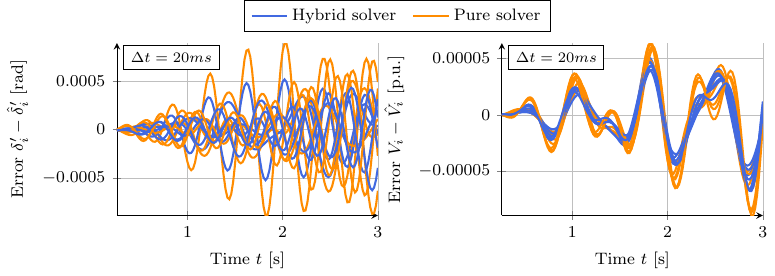}
\caption{Error evolutions for a load disturbance at bus 45 over a three-second simulation with a time step $\Delta t = 20 \, \textrm{ms}$. In the left plot, each line depicts the error evolution of the rotor angle of each generator. In the right plot, each line depicts the error evolution of the voltage magnitude of each bus in the system. The blue and orange colors represent the hybrid and pure solver, respectively.}\label{fig:57bussystem}
\end{figure}

We simulate the IEEE 57-bus system after a load disturbance at a random bus with the pure and hybrid solvers. The simulations are then compared to the true trajectory, which is simulated with a very small time step. \Cref{fig:57bussystem} shows the error evolution of the rotor angles for the seven machines and the voltage magnitudes for the 57 buses. As shown for the 9-bus system, the dynamics of the machines captured by PINNs are simulated significantly more accurately, boosting their accuracy by 51 \%. The accuracy boost on those four machines single-handedly improves the performance of the overall simulation, showing a 24 \% boost in the presented simulation, allowing the simulation to take larger steps without losing accuracy.

\section{Discussion} \label{sec5:discussion}
The results in the previous Sec.~\ref{sec4:numstudy} illustrate how replacing conventional numerical methods with NN-based methods in the established transient simulation framework can increase the simulation's accuracy and time step size; thus speeding up current simulation algorithms. This integration comes at the cost of training the PINN beforehand. However, this up-front training cost should not be seen as a disadvantage but rather as a method's flexibility to offload computational time from the application side. By accurately training the NN-based method beforehand, we achieve a tailored numerical scheme for the specific component. Thus, the simulation becomes more accurate and allows for larger time step sizes that would not be possible with the typical Runge-Kutta schemes. In other words, we incorporate knowledge into the method to provide more accurate simulation results. One could envision the future to have a library of pre-trained components, similar to the library of models that current simulators have, which the user could then ``mix and match'' in a conventional simulation environment and perform simulations at significantly higher speed.

The NN-based methods with the formulation we introduce in this paper can be easily and accurately trained. Thus, the limiting factors to further increase the time step sizes are the propagated errors from other components integrated with classical schemes and the errors incurred from the difference between the assumed and true $y(x)$ profile. We envision further developments in improving the $y(x)$ profile and in the methods of output verification.



\section{Conclusions} \label{sec6:conclusions}
This paper presents the first natural step to modularly integrate Physics-Informed Neural Networks (PINNs) into transient simulation workflows. We develop a PINN formulation that learns the dynamics of individual components considering the system's interaction. This novel formulation enables PINNs as an alternative to conventional methods, providing more accurate results over larger time step sizes. We show that replacing the dynamics of fast-evolving components with PINNs significantly reduces the simulation errors, allowing for larger time steps at the same accuracy. This unlocks a better synergy between time-domain simulations and the emerging machine learning methods, where instead of changing the current simulation paradigm, we leverage PINNs to improve its performance. Besides introducing the methodology for the IEEE 9- and 57-bus systems, we make a first version of the developed toolbox available online \cite{git} to foster the adoption of PINNs in the power systems community. Future work will capture more dynamic components, more degrees of freedom in the variables' profiles, and include electromagnetic phenomena.


\section*{Acknowledgement}
I. Ventura Nadal and S. Chatzivasileiadis were supported by the ERC Starting Grant VeriPhIED, funded by the European Research Council, Grant Agreement 949899. J. Stiasny was supported by the joint PhD programme between TU Delft and AIT.

\end{document}